\documentclass[prl,aps,twocolumn,amsmath,amssymb]{revtex4}
\usepackage{graphicx}
\usepackage{bm}
\begin{document}

\title{Dissipation in circuit quantum electrodynamics: \\
lasing and cooling of a low-frequency oscillator}

\author{Julian Hauss$^{1,2}$, Arkady Fedorov$^{1,3}$,
Stephan Andr\'e$^1$, Valentina Brosco$^1$, Carsten Hutter$^{1,4}$,\\
Robin Kothari$^{1,5}$, Sunil Yeshwanth$^{1,6}$,
Alexander Shnirman$^{1,7}$, and Gerd Sch\"on$^1$}

\affiliation{$^1$ Institut f\"{u}r Theoretische
Festk\"{o}rperphysik and DFG-Center for Functional Nanostructures
(CFN), Universit\"{a}t Karlsruhe, D-76128 Karlsruhe, Germany}
\affiliation{$^2$ Lichttechnisches Institut, Universit\"{a}t
Karlsruhe, D-76128 Karlsruhe, Germany}
\affiliation{$^3$Kavli Institute of Nanoscience, Delft University of
Technology, 2600 GA Delft, The Netherlands}
\affiliation{$^4$Department of Physics, Stockholm University,
AlbaNova University Center, SE - 106 91 Stockholm, Sweden}
\affiliation{$^{5}$Department of Physics, Indian Institute of
Technology Bombay, Mumbai 400076, India}
\affiliation{$^{6}$Department of Physics, Indian Institute of
Technology, Kanpur 208016, India}
\affiliation{$^7$ Institut f\"{u}r Theorie der Kondensierten Materie,
Universit\"{a}t Karlsruhe, D-76128 Karlsruhe, Germany}

\begin{abstract}
Superconducting qubits coupled to electric or nanomechanical
resonators display  effects previously studied in quantum
electrodynamics (QED) and extensions thereof. Here we study
a driven qubit coupled to a low-frequency tank circuit
with particular emphasis on the role of dissipation.
When the qubit is driven to perform Rabi oscillations, with Rabi
frequency  in resonance with the oscillator, the latter can be
driven far from equilibrium.
Blue detuned driving leads to a population inversion in the qubit
and lasing behavior of the oscillator (``single-atom laser"). For red detuning the qubit
cools the oscillator. This behavior persists at the symmetry point where
the qubit-oscillator coupling is quadratic
and decoherence effects are minimized. Here
the system realizes a ``single-atom-two-photon laser".
\end{abstract}

\maketitle

\section{Introduction}

\begin{figure}
\includegraphics[width=7cm]
{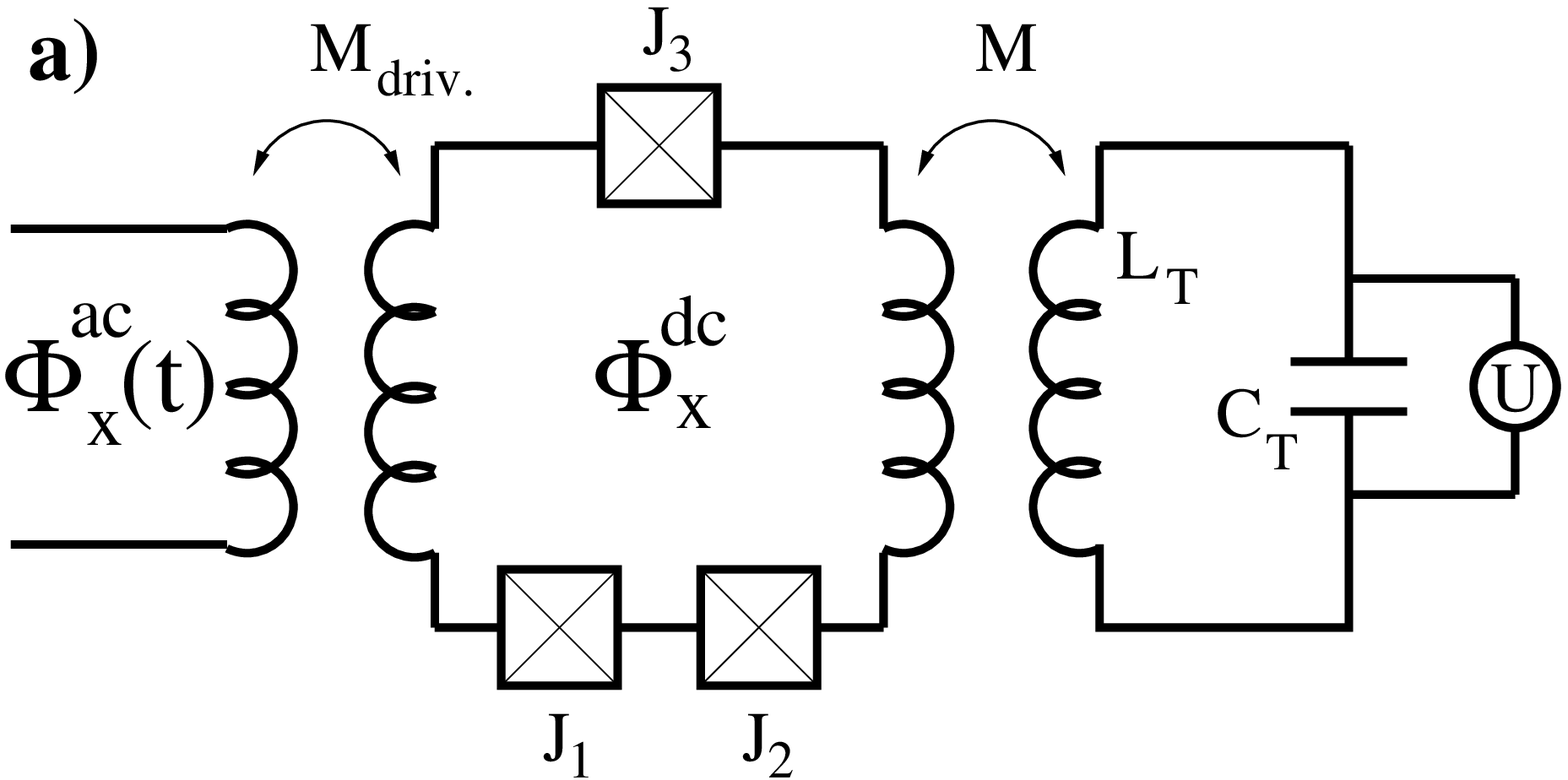}
\vskip 5mm
\includegraphics[width=5.5cm]
{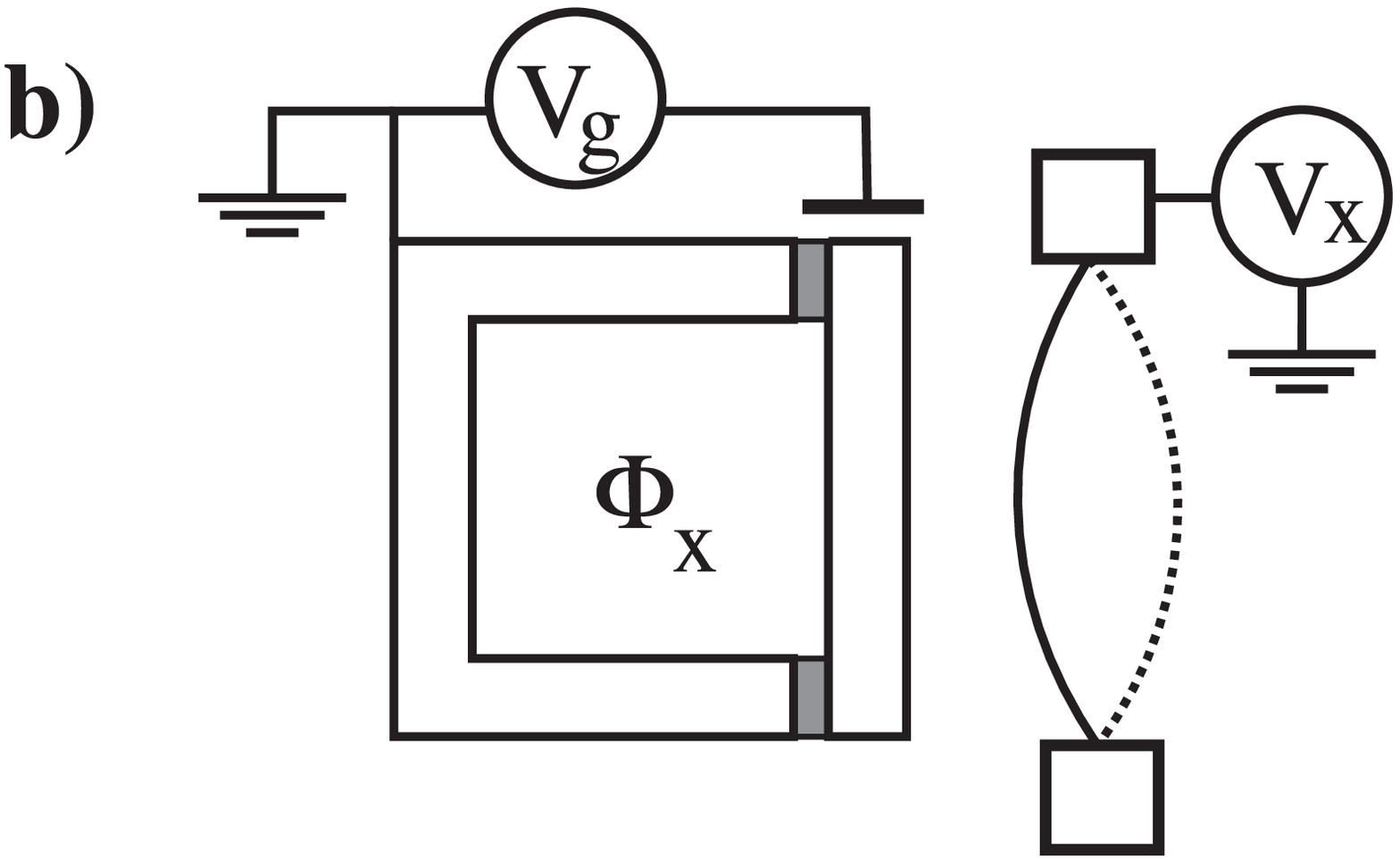}
\caption{a) In the setup of Ref.~\cite{Jena_Rabi} an externally driven
three-junction flux qubit is coupled inductively to an LC
oscillator. b) A charge qubit is coupled to a mechanical
resonator.} \label{fig:system}
\end{figure}

Recent experiments on quantum state engineering with superconducting
circuits realized concepts originally introduced in the field of
quantum optics, as well as extensions thereof, e.g., to the regime
of strong coupling~\cite
{Jena_Rabi,Wallraff_CQED,Chiorescu_CQED,wallraff-2005-95,
PhysRevLett.96.127006,naik-2006-443,grajcar-2007,deppe-2008}, and
prompted substantial theoretical
activities~\cite{Buisson_QED,blais-2004-69,liu-2004-67,
martin-2004-69,moon:140504,
mariantoni-2005,liu-2006-74,wallquist-2006,xue-2006,
Hauss_PRL2008,ashhab-2008}. Josephson qubits play the role of
two-level atoms while electric or nanomechanical oscillators play
the role of the quantized radiation field. In most QED or circuit
QED experiments the atom or qubit transition frequency is near
resonance with the oscillator. In contrast, in the experiments of
Refs.~\cite{Jena_Rabi}, with setup shown in Fig.~\ref{fig:system}a),
the qubit is coupled to a slow LC oscillator with frequency
($\omega_T/2\pi\sim$ MHz) much lower than the qubit's level
splitting ($ \Delta E/2\pi \hbar \sim 10$ GHz). The idea of this
experiment  is to drive the qubit to perform Rabi oscillations with
Rabi frequency in resonance with the oscillator, $\Omega_{R} \approx
\omega_{T}$. In this situation the qubit should drive the oscillator
and increase its oscillation amplitude. When the qubit driving
frequency is blue detuned, the driving creates a population
inversion of the qubit, and the system exhibits lasing behavior
(``single-atom laser"); for red detuning the qubit cools the
oscillator~\cite{Hauss_PRL2008}. A similar strategy for cooling of a
nanomechanical resonator via a Cooper pair box qubit has been
recently suggested in Ref.~\cite{wallquist-2008}. The analysis of
the driven circuit QED system shows that these properties depend
strongly on relaxation and decoherence effects in the qubit.

In the experiments of Ref.~\cite{Jena_Rabi} an enhancement of
the oscillator due to the driving was observed.  However, first attempts to
explain the effect theoretically did not resolve several issues~\cite
{Smirnov,greenberg-2005-72}. In the experiments, in order to minimize
decoherence effects, the Josephson flux
qubit was biased near the flux degeneracy point. At this symmetry
point also the coupling to the oscillator is tuned to zero, and the
enhancement should vanish. Uncontrolled small deviations from
the symmetry point might lead to the observed effect~\cite{Jena_Rabi,Smirnov}, but this explanation has not been supported by experiments. Here we
explore an alternative, namely that a quadratic
coupling to the oscillator near the resonance condition  $\Omega_{R}
\approx 2 \omega_{T}$ is responsible for the observed enhancement.
In the following we will consider both
linear as well as quadratic coupling, which dominate away
from the symmetry point and at this point, respectively.
The second unresolved problem is the magnitude of the effect. The
experiments~\cite{Jena_Rabi} showed an increase by a factor $4- 5$ in
the amplitude, i.e.,
$16 - 25$ in the number of oscillator quanta. The
theory of Ref.~\cite{Smirnov}, valid in the perturbative regime,
predicts a much weaker effect.
We obtain a strong effect as follows~\cite{Hauss_PRL2008}: due to a
detuning of the qubit driving a population inversion is
created at the Rabi frequency, and the system becomes a
``single-atom laser'' at the resonance $\Omega_{R} \approx \omega_{T}$,
or a ``single-atom-two-photon laser'' for
$\Omega_{R} \approx 2 \omega_{T}$~\cite{PhysRevA.46.5944,
mckeever-2003-425}.
In both cases the lasing threshold is reached for realistic system
parameters, and the number of quanta in the oscillator is increased
considerably.

A related situation, called ``dressed-state lasing'', had been studied
before in quantum optics~\cite{Zakrewski,PhysRev.188.1969}. The present scenario
differs from that one in so far as the resonator modes are coupled to
the low-frequency Rabi oscillations
rather than to the high-frequency Mollow
transitions. The Rabi frequency can be
readily tuned to resonance with the oscillator, which should facilitate
reaching the lasing threshold and a proper lasing state.
A similar idea has been explored in Ref.~\cite{Plenio} in connection with coupling of atoms.

In experiments with the same setup as shown in
Fig.~\ref{fig:system}a) but in a different parameter regime
the mechanisms of Sisyphus cooling and
amplification has recently been demonstrated~\cite{grajcar-2007}.
Due to the resonant high-frequency driving of the qubit, depending on the detuning, the oscillator is either cooled or amplified with a tendency towards lasing.
The Sisyphus mechanism is most efficient when the relaxation rate of the
qubit is close to the oscillator's frequency. In contrast, in
the present paper we concentrate on the ``resolved sub-band" regime where the
dissipative transition
rates of the qubits are much lower than the oscillator's frequency.

Also in situations where the qubit, e.g., a Josephson charge
qubit, is coupled to a nano-mechanical oscillator
(Fig.~\ref{fig:system}b) it either cools or amplifies the oscillator.
On one hand, this may constitute an important tool on the way to ground state cooling.
On the other hand, this setup provides a realization of what is called a SASER~\cite{SASER}.

Lasing and cooling of the oscillator has also been observed in a
slightly different setup, when the
$ac$-driven qubit is replaced by a driven superconducting
single-electron transistor biased near the Josephson quasiparticle
cycle~\cite{Fulton1989,Geerligs,Maassen1991b}.
When the SSET is coupled to a nanomechanical or electric oscillator
it can be used to either cool the oscillator
~\cite{blanter-2004-93,blencowe-2005-7,clerk-2005-7,bennett-2006-74,
usmani-2007,rodrigues-2006} or to produce laser-like behavior.
The latter has recently been observed in experiments~\cite{Astafiev07}.

\section{The system}

\subsection{The Hamiltonian}

The systems to be considered are shown in Fig.~\ref{fig:system}.
A qubit is coupled to an oscillator and
driven to perform Rabi oscillations.
To be specific we first analyze the flux qubit coupled to an electric oscillator
(Fig.~\ref{fig:system}a) with Hamiltonian
\begin{eqnarray}
\label{eq:Hamiltonian} H & = &
-\frac{1}{2}\,\epsilon\left(\Phi_{x}^{dc}\right){\sigma}_{z}
-\frac{1}{2}\, \Delta {\sigma}_{x} -
\hbar\Omega_{R0}\cos(\omega_{d}t)\, {\sigma}_{z}\nonumber\\
   &  &+\ \hbar\omega_{T}\,{a}^{\dagger}
{a} \ +\ g\, {\sigma}_{z}\left({a}+{a}^{\dagger}\right)\ .
\end{eqnarray}
The first two terms describe the qubit, with Pauli matrices
${\sigma}_{x,z}$ operating in the flux basis of the qubit. The
energy bias between the flux states $\epsilon(\Phi_{x}^{dc})$ is
controlled by an external DC magnetic flux,
   and $\Delta$ is the tunneling amplitude between the basis
states. The resulting level spacing
$\Delta E \equiv \sqrt{\epsilon^2 + \Delta^2}$ typically lies in the
range of several GHz.
The third term accounts for the driving of the qubit by an
applied AC magnetic flux with amplitude $\Omega_{R0}$ and
frequency $\omega_{d}$. The last two terms describe the oscillator with
frequency $\omega_T=1/\sqrt{L_T C_T}$', which for the experiments of
of Ref.~\cite{Jena_Rabi} lies in the range of several 10 MHz,
as well as the qubit-oscillator
interaction. We estimate the coupling constant
$g\approx M I_p I_{T,0}$ to be of the order of 10 MHz.
Here $M$ is the mutual inductance,
$I_p$ the magnitude of the persistent current in the qubit, and
$I_{T,0}=\sqrt{\hbar\omega_{T}/2L_{T}}$ the amplitude of the
vacuum fluctuation of the current in the LC oscillator.

After transformation to the eigenbasis of the qubit, which is the
natural
basis for the description of the dissipation, the Hamiltonian reads
\begin{eqnarray}
\label{eq:H_eigenbasis} {H} & = & -\frac{1}{2}\, \Delta
E\,{\sigma}_{z}- \,\hbar\Omega_{R0}\cos\left(\omega_{d}t\right)
   \left(\sin\zeta\,{\sigma}_{z}
-\cos\zeta\,{\sigma}_{x}\right)\nonumber\\
   & + & \hbar\omega_{T}\,{a}^{\dagger}{a}\
+\, g\left(\sin\zeta\,{\sigma}_{z}-
   \cos\zeta\,{\sigma}_{x}\right)\left({a}
+{a}^{\dagger}\right)\ ,
\end{eqnarray}
with $\tan\zeta=\epsilon/\Delta$ and
$\Delta E \equiv \sqrt{\epsilon^2 + \Delta^2}$.

Because of the large difference of the energy scales between the
qubit and the oscillator, $\Delta E \gg \hbar \omega_{T}$, it is
tempting, in the spirit of the usual rotating wave
approximation (RWA), to drop the transverse coupling term
$-g\cos\zeta\,{\sigma}_{x}\left({a}
+{a}^{\dagger}\right)$ of Eq.~(\ref{eq:H_eigenbasis}).
However, near the symmetry point (where $\sin\zeta = 0$)
the longitudinal coupling
is weak. Therefore, we retain the transverse coupling, but transform
it by employing a Schrieffer-Wolff transformation,
${U}_{S}=\exp\left(i{S}\right)$, with generator ${S}=(g/\Delta
E)\cos \zeta\,\left({a}+{a}^{\dagger}\right)\,{\sigma}_{y}$, into
a second-order longitudinal coupling. On the other hand, since
$\hbar \omega_{d}\sim \Delta E$, we can drop within RWA the
longitudinal driving term
$-\hbar\Omega_{R0}\cos\left(\omega_{d}t\right)\sin\zeta\,{\sigma}_{z}$.
The Hamiltonian then reads
\begin{eqnarray}
\label{eq:H_RWA} {H} & = & -\frac{1}{2}\,\Delta
E\,{\sigma}_{z}
+\hbar\Omega_{R0}\cos\left(\omega_{d}t\right)
\cos\zeta\,{\sigma}_{x}
+\hbar\omega_{T}\, {a}^{\dagger}{a}\nonumber\\
&+&g\sin\zeta\,{\sigma}_{z}\left({a}+{a}^{\dagger}\right)
-\frac{g^{2}}{\Delta E}\cos^{2}\zeta\,{\sigma}_{z}
\left({a}+{a}^{\dagger}\right)^{2}\, .
   \end{eqnarray}
A further unitary transformation with
${U}_{R}=\exp\left(-i\omega_{d}{\sigma}_{z}t/2\right)$
brings the Hamiltonian to the rotating frame,
$\tilde H \equiv {U}_{R}^{\phantom{\dag}}H{U}_{R}^{\dag}+ i\hbar {\dot
U}_{R}^{\phantom{\dag}}{U}_{R}^{\dag}$. We obtain
\begin{eqnarray}
\label{eq:Hamilton in RF} \tilde H & =& \frac{1}{2}\,
\hbar\delta\omega\, {\sigma}_{z} +\frac{1}{2}\,
\hbar\Omega_{R0}\, \cos\zeta\, {\sigma}_{x}+\hbar\omega_{T}\,{a}^
{\dagger}{a}\nonumber \\
   &+& g\sin\zeta\,{\sigma}_{z}\left({a}+{a}^{\dagger}\right)
-\frac{g^{2}}{\Delta E}\cos^{2}\zeta\,{\sigma}_{z}\left({a}+{a}^
{\dagger}\right)^{2}.\nonumber\\
\end{eqnarray}
Here $\delta\omega \equiv \omega_{d}-\Delta E/\hbar$ is the detuning.
After diagonalization of the qubit terms we obtain
\begin{eqnarray}
\label{eq:Hamilton in rot Bezugssyst mit quadr Koppl}
\tilde H & = &
\frac{1}{2}\,\hbar\Omega_{R}\,{\sigma}_{z}+\hbar\omega_{T}\,
{a}^{\dagger}{a}
\nonumber\\
   & + & g\sin\zeta\,\left[\, \sin\beta\,{\sigma}_{z}-
   \cos\beta\,{\sigma}_{x}\right]
\left({a}+{a}^{\dagger}\right)\nonumber \\
   & - & \frac{g^{2}}{\Delta E}\, \cos^2\zeta\,\left[\,
   \sin\beta\,{\sigma}_{z}
-\cos\beta\,{\sigma}_{x}\right]\left({a}
+{a}^{\dagger}\right)^{2}\, .
\end{eqnarray}
Here
$\Omega_{R}=\sqrt{\Omega_{R0}^{2}\cos^2\zeta+\delta\omega^{2}} $
and $\tan\beta=\delta\omega/(\Omega_{R0}\cos\zeta)$.

Finally we employ a second RWA. While the first one dropped terms
oscillating with frequencies of order $\Delta E/\hbar$, the second
one assumes the Rabi frequency $\Omega_{R}$ and the oscillator
frequency $\omega_{T}$ to be fast. In the interaction
representation with respect to the non-interacting Hamiltonian,
$\tilde H_0=(\hbar\Omega_R/2){\sigma}_z+\hbar\omega_T\,
{a}^{\dagger}{a}$, we then obtain
\begin{eqnarray}
\tilde H_{I} &=&g_1\left(a^{\dagger}\sigma_-e^{-i(\Omega_R-\omega_T)
t}+h.c.\right)
\label{V_I}\nonumber \\
&+&g_2\left(a^{\dagger2}\sigma_-e^{-i(\Omega_R-2\omega_T)
t}+h.c.\right)\nonumber\\
&+&g_3\left(a^{\dagger}a +a a^{\dagger} \right)\sigma_z\ .
\end{eqnarray}
We kept both single-photon and two-photon interactions with
$g_1=g\sin\zeta \cos \beta$ and $g_2=(g^2/\Delta E)\cos^2\zeta \cos
\beta$, although within  RWA only one of them survives: the
single-photon term for $\Omega_{R} \sim \omega_{T}$, or the
two-photon term for $\Omega_{R}\sim 2\omega_{T}$. The last term of
(\ref{V_I}) with $g_3 = -(g^2/\Delta E)\cos^2\zeta \sin \beta$
is the $ac$-Stark effect, causing a qubit state dependent
frequency shift of the oscillator~\cite{greenberg-2002-66}.
In what follows
we will assume that the qubit is kept near the symmetry point,
i.e., $\epsilon\ll\Delta$ and $\cos\zeta\simeq1$.

\subsection{Transition rates in the rotation frame}

The transformation to ``dressed states" in the rotating frame modifies the relaxation,
excitation and decoherence rates as compared to the standard
results.
To illustrate these effects and justify the treatment of the
dissipation in latter sections we first consider a driven qubit (ignoring
the coupling to the oscillator) coupled to a bath observable $\hat
X_{B}$,
\begin{eqnarray}
\label{eq:H_eigenbasis_NO} {H} & = & -\frac{1}{2}\, \Delta
E\,{\sigma}_{z}+\,\hbar\Omega_{R0}\,
\cos\left(\omega_{d}t\right)\,{\sigma}_{x}\nonumber\\
   & - &\frac{1}{2} \left(b_x \sigma_x +
   b_y \sigma_y + b_z \sigma_z\right)\hat X_{B}+
H_{\rm bath}\ .
\end{eqnarray}

In the absence of driving, $\Omega_{R0}=0$,
and for regular power spectra of the fluctuating bath observables
we can proceed using Golden rule type arguments.
The transverse noise is responsible for the relaxation and excitation
with rates
\begin{eqnarray} \label{eq:Gamma_bar}
\Gamma_{\downarrow} & = & \frac{|b_\perp|^2}{4\hbar^2}
\langle\hat{X}_{B}^{2}(\omega=\Delta E)\rangle \nonumber \\
\Gamma_{\uparrow} & = & \frac{|b_\perp|^2}{4\hbar^2}
\langle\hat{X}_{B}^{2}(\omega=-\Delta E)\rangle \, ,
\end{eqnarray}
while longitudinal noise produces a pure dephasing with rate~\cite
{Saclay_Karlsruhe}
\begin{eqnarray} \label{eq:Gamma_*}
\Gamma_{\varphi}^* & = &  \frac{|b_z|^2}{2\hbar^2} S_{X}(\omega =0) \ .
\end{eqnarray}
Here $b_\perp \equiv b_x + i b_y$, and we introduced the ordered
correlation function
$
\langle\hat{X}_{B}^{2}(\omega)\rangle\equiv\int dt\ e^{i\omega
t}\langle\hat{X}_{B}(t)\hat{X}_{B}(0)\rangle\ ,
$
as well as the power spectrum, i.e., the symmetrized correlation
function, $S_{X}(\omega)\equiv
(\langle\hat{X}_{B}^{2}(\omega)\rangle+\langle\hat{X}_{B}^{2}(-\omega)
\rangle)/2$.
The rates (\ref{eq:Gamma_bar}) and (\ref{eq:Gamma_*}) also define the relaxation rate $1/T_1=
\Gamma_1= \Gamma_{\downarrow}
+\Gamma_{\uparrow} $ and the total dephasing rate $1/T_2 =\Gamma_
\varphi= \Gamma_1/2  + \Gamma_{\varphi}^*$ which appear in the Bloch
equations.

To account for the driving with frequency $\omega_d$ it is convenient
to transform to the rotating frame via a unitary transformation $U_R =
\exp{(-i\omega_d\sigma_z t/2)}$. Within RWA the transformed Hamiltonian
reduces to
\begin{eqnarray}
\label{eq:H_Rotating}
\tilde H &= &\frac{1}{2}\hbar\left[\Omega_{R0}{\sigma}_{x}+ \delta
\omega{\sigma}
_{z}\right]\\
&-&\frac{1}{2}\left[b_z{\sigma}_{z}+ b_\perp e^{i\omega_d t}\sigma_-
+ b_\perp^* e^{-i\omega_d t}\sigma_+
\right]\hat X_{B} + H_{\rm
bath}\ ,\nonumber
\end{eqnarray}
where $b_\perp \equiv b_x + i b_y$, and the detuning is
$\delta\omega\equiv\omega_{d}-\Delta E/\hbar$.
The RWA cannot be used in the second line of
(\ref{eq:H_Rotating}) since the fluctuations
$\hat X_{B}$ contain potentially all frequencies,
including those of order $\pm \omega_d$ which can compensate  fast
oscillations.

Diagonalizing the first line of (\ref{eq:H_Rotating})
one obtains
\begin{eqnarray}
\label{eq:H_Rotating_diag}
&&\tilde H = \frac{1}{2}\hbar \Omega_{R}\,\sigma_z+H_{\rm bath}
\nonumber \\
&&-\left[\frac{\sin\beta}{2}\, b_{z}+\frac{\cos\beta}{4} \,(b_\perp^*\,
e^{-i\omega_d t} + b_\perp\, e^{i\omega_d t}) \right]\,\sigma_z \hat
X_{B}
\nonumber\\
&&-\Big\{\Big[\frac{(\sin\beta+1)}{4}\,b_\perp^*\, e^{-i\omega_d t} +
\frac{(\sin\beta-1)}{4}\,b_\perp \,e^{i\omega_d t} \nonumber\\
&&-\frac{\cos\beta}{2} \,b_z
\Big]\,\sigma_+\, \hat X_{B} + h.c.\Big\}\ ,
\end{eqnarray}
with
$\Omega_{R}=\sqrt{\Omega_{R0}^{2}+\delta\omega^{2}} $ and  $\tan\beta=
\delta\omega/\Omega_{R0}$.

From here Golden-rule arguments yield the relaxation and excitation
rates in the rotating frame~\cite{Saclay_Karlsruhe}
\begin{eqnarray} \label{eq:Gamma_down}
\tilde\Gamma_{\downarrow} & \approx &
\frac{b_{z}^{2}\cos^{2}\beta}{4\hbar^{2}}
\langle\hat{X}_{B}^{2}(\omega=\Omega_{R})\rangle \\
&+&\frac{|b_\perp|^2}{16\hbar^{2}}\left(1-\sin\beta\right)^2
\langle\hat{X}_{B}^{2}(\omega=\omega_{d}+\Omega_{R})\rangle\nonumber \\
& + & \frac{|b_\perp|^2}{16\hbar^{2}}
\left(1+\sin\beta\right)^2
\langle\hat{X}_{B}^{2}(\omega=-\omega_{d}+\Omega_{R})\rangle\
\nonumber
\end{eqnarray}
and
\begin{eqnarray}
\label{eq:Gamma_up}
\tilde\Gamma_{\uparrow} & \approx &
\frac{b_{z}^{2}\cos^{2}\beta}{4\hbar^{2}}
\langle\hat{X}_{B}^{2}(\omega=-\Omega_{R})\rangle\\
& + & \frac{|b_\perp|^2}{16\hbar^{2}}
\left(1-\sin\beta\right)^2
\langle\hat{X}_{B}^{2}(\omega=-\omega_{d}-\Omega_{R})\rangle\nonumber \\
& + & \frac{|b_\perp|^2}{16\hbar^{2}}
\left(1+\sin\beta\right)^2
\langle\hat{X}_{B}^{2}(\omega=\omega_{d}-\Omega_{R})\rangle\ ,\nonumber
\end{eqnarray}
as well as the "pure" dephasing rate
\begin{eqnarray}
\label{eq:Gamma_puredeph}
\tilde\Gamma_{\varphi}^*  \approx
\frac{b_{z}^{2}\sin^{2}\beta}{2\hbar^{2}}S_
{X}(\omega=0)
+\frac{\cos^{2}\beta}{4\hbar^{2}}
|b_\perp|^2 S_{X}(\omega=\omega_{d})\
.\nonumber\\\end{eqnarray}
We note the effect of the frequency mixing, and due to the
diagonalization the effects of longitudinal and transverse noise on
relaxation and decoherence get mixed. In addition we note that
the rates also depend on the fluctuations' power spectrum at the
Rabi frequency.

For a sufficiently regular power spectrum of fluctuations with
$\omega \approx \pm \Delta E/\hbar$, we can ignore the effect of detuning
and the small
shifts by $\pm \Omega_R$ as compared
to the high frequency $\omega_{d}\approx \Delta E/\hbar$.
We further assume that $\Omega_{R}\ll k_B T /\hbar$.
In this case we find the simple relations
\begin{eqnarray}
\label{mod}
\tilde \Gamma_\uparrow & = &
\frac{(1+\sin \beta)^2}{4}\, \Gamma_\downarrow
+ \frac{(1-\sin \beta)^2}{4}\, \Gamma_\uparrow
+ \frac{1}{2}\cos^2 \beta \, \Gamma_\nu \ , \nonumber \\
\tilde \Gamma_\downarrow & = &
\frac{(1-\sin \beta)^2}{4}\, \Gamma_\downarrow
+ \frac{(1+\sin \beta)^2}{4}\, \Gamma_\uparrow
+ \frac{1}{2}\cos^2 \beta \, \Gamma_\nu \ , \nonumber \\
\tilde\Gamma_{\varphi}^*   &=& \sin^2\beta\, \Gamma_{\varphi}^*+
\frac{\cos^{2}\beta}{2}(\Gamma_\downarrow+\Gamma_\uparrow)\ ,
\end{eqnarray}
where the rates in the lab frame are given by Eq.~(\ref{eq:Gamma_bar})
and the new rate,
$\Gamma_\nu  \equiv  \frac{1}{2} \,b_{z}^{2} \,S_{X}(\Omega_R)$,
depends on the power spectrum at the Rabi frequency.

To proceed we concentrate on the
regime relevant for our system.  Near the symmetry point of the qubit we
have $b_z \approx 0$. At low temperatures, $k_{\rm B}T \ll \Delta E
\approx \hbar\omega_d$, we can neglect $\Gamma_\uparrow$ as it is
exponentially small.
Thus we are left with
\begin{eqnarray} \label{eq:Gamma_down_res}
\tilde\Gamma_{\downarrow/\uparrow} \approx
\frac{\left(1\mp\sin\beta\right)^2}{4}
\,\Gamma_0\quad  ,\quad \tilde\Gamma_{\varphi}^*   \approx
\frac{\cos^{2}\beta}{2}\Gamma_0¾\ ,
\end{eqnarray}
where
\begin{equation}\label{eq:Gamma_0}
\Gamma_0\equiv  \frac{|b_\perp|^2}{4\hbar^{2}}
\langle\hat{X}_{B}^{2}(\omega=\Delta E/\hbar)\rangle \approx \Gamma_
\downarrow\ .
\end{equation}

The ratio of up- and down-transitions depends on the detuning and can
be expressed by an effective temperature. Right on
resonance, where $\beta=0$, we have
$\tilde\Gamma_{\uparrow}=\tilde\Gamma_{\downarrow}$, corresponding to
infinite
temperature or a classical drive.  For
``blue" detuning, $\beta > 0$, we find $\tilde\Gamma_{\uparrow} >
\tilde\Gamma_{\downarrow}$, i.e., {\it negative temperature}.
This leads to a population inversion of the qubit, which is the basis
for the lasing behavior which will be described below.

In a more careful analysis, not making
use of the ``white noise" approximation, we obtain for $\beta =0$
\begin{equation}
\frac{\tilde\Gamma_\downarrow}{\tilde\Gamma_\uparrow} =
\frac{\langle\hat{X}_{B}^{2}(\omega=\omega_{d}+\Omega_{R})\rangle}
{\langle\hat{X}_{B}^{2}(\omega=\omega_{d}-\Omega_{R})\rangle} \, .
\end{equation}
For instance, for Ohmic noise and low bath temperature this reduces to $\tilde\Gamma_\downarrow/\tilde\Gamma_\uparrow
\sim 1 +
2\Omega_{R}/\omega_d$. This corresponds to an effective
temperature of order $2\hbar\omega_d/k_{\rm B}\sim 2\Delta
E/k_{\rm B}$, which by assumption is high but finite. The infinite
temperature threshold is crossed toward negative temperatures
at weak blue detuning when the condition
\begin{equation}
\frac{(1+\sin\beta)^2}{(1-\sin\beta)^2} \sim 1+\frac{2\Omega_{R}}
{\omega_d}\
\end{equation}
is satisfied.

\begin{figure}
\includegraphics[width=0.8\columnwidth]
{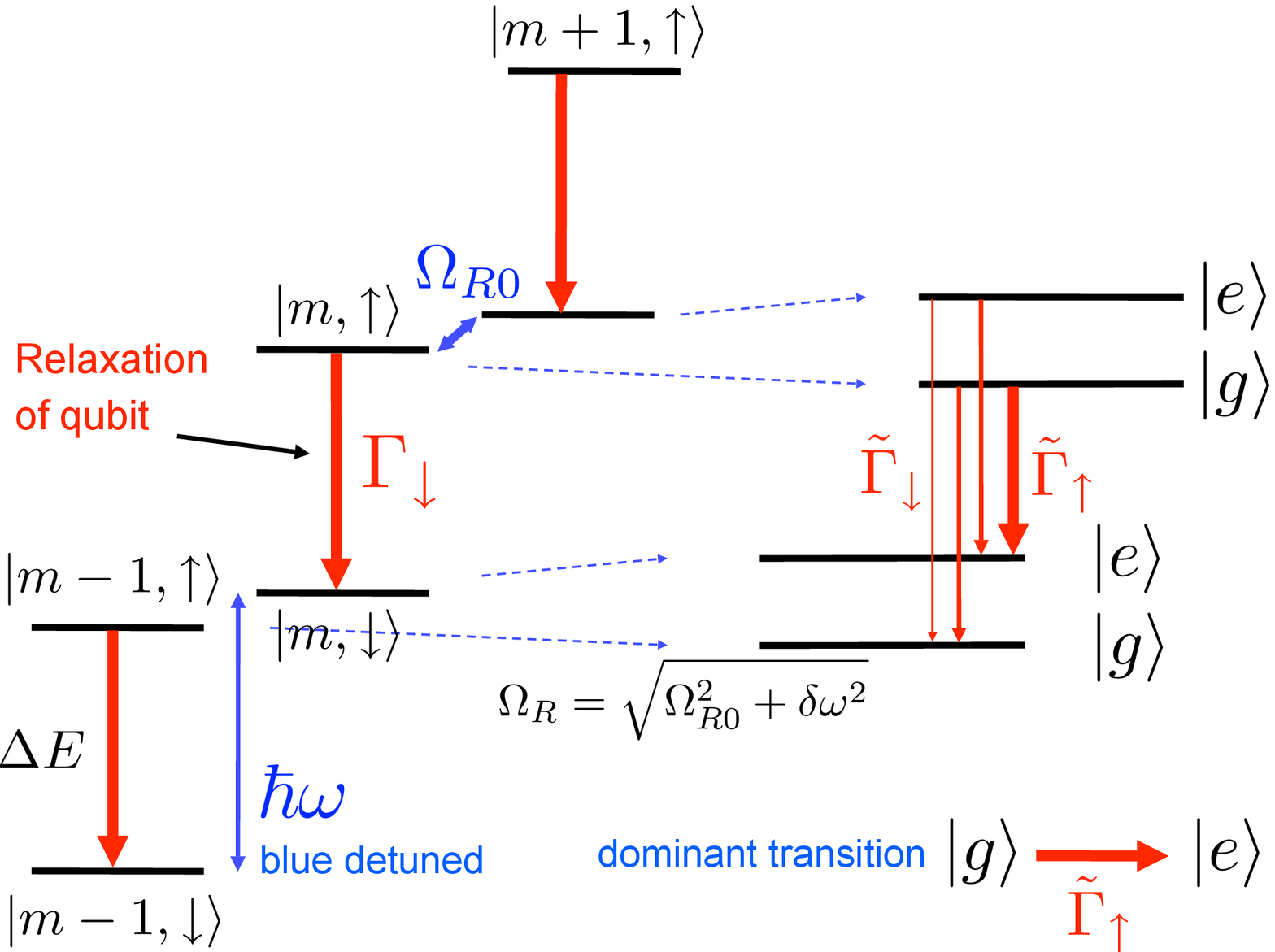} \caption{Dressed states of a driven qubit near resonance.
Here $m$ is the number of photons of the driving field,
which is assumed to be quantized.
}
\label{fig:mollow}
\end{figure}

To illustrate how the population inversion is created for blue
detuning we show in Fig.~\ref{fig:mollow} the level structure, i.e., the formation of dressed states, of a near-resonantly driven qubit.
For the purpose of this explanation the driving field is quantized. This level
structure was described first by Mollow~\cite{PhysRev.188.1969}.
The picture also illustrates how for blue detuning
a pure relaxation process, $\Gamma_\downarrow=\Gamma_0$, in the laboratory frame predominantly leads to an excitation process, $\tilde\Gamma_\uparrow$, in the rotating frame.

\subsection{The Liouville equation in the rotating frame}

Within the approximation leading to the rates (\ref{mod})
the Liouville equation governing the dynamics of the
density matrix  in the rotating frame can presented in a
simple Lindblad form.
Start in the lab frame the dissipation is accounted
for by two damping terms,
\begin{equation}
\dot \rho=-\frac{i}{\hbar}\left[H,
\rho\right]+L_Q\,\rho+L_R\,\rho\ .
\end{equation}
The qubit's dissipation is described by
\begin{eqnarray}\label{L_Q^R}
L_Q\rho &=&\frac{\Gamma_{\downarrow}}{2}\left(2 \sigma_- \rho
\sigma_+
-\rho\sigma_+\sigma_- -\sigma_+\sigma_-\rho \right)\nonumber\\
&+&\frac{\Gamma_{\uparrow}}{2}\left(2 \sigma_+ \rho\sigma_-
-\rho\sigma_-\sigma_+ - \sigma_-\sigma_+\rho \right)\nonumber\\
&+&\frac{\Gamma_{\varphi}^*}{2} \left(\sigma_z\rho\sigma_z -
\rho \right)\ ,
\end{eqnarray}
with rates given by (\ref{eq:Gamma_bar}) and (\ref{eq:Gamma_*}).
The resonator damping, with strength parametrized by $\kappa$,
can be  written as~\cite{Gardiner}
\begin{eqnarray}\label{damp}
L_R \rho=&&\frac{\kappa}{2}\left(N_{\rm th}+1\right)\left( 2a \rho a^
{\dagger}
-a^{\dagger} a\rho -\rho a^{\dagger}a \right)\nonumber\\
&&+\frac{\kappa}{2}N_{\rm th} \left( 2 a^{\dagger} \rho a-a
a^{\dagger}\rho -\rho aa^{\dagger} \right)\ .
\end{eqnarray}
Here
$N_{\rm th}=1/\left[\exp(\hbar\omega_T/k_B T)-1\right]$  is the thermal
distribution function of photons in the resonator.

The transformations to the rotating frame and RWA described above
transform the Liouville equation to a new form, which in our
approximation
again has a Lindblad form. In the interaction representation it is
\begin{equation}
\label{eq:Master_Equation}
\dot{\tilde\rho}=-\frac{i}{\hbar}\left[\tilde{H}_I,\tilde{\rho}
\right]
+\tilde{L}_Q \, \tilde{\rho}+L_R \, \tilde{\rho} \ ,
\end{equation}
where
\begin{eqnarray}\label{L_Q^R_rotating}
\tilde{L}_Q \tilde\rho &=&\frac{\tilde\Gamma_{\downarrow}}{2}\left(2
\sigma_-
\tilde\rho
\sigma_+
-\tilde\rho\sigma_+\sigma_- -\sigma_+\sigma_-\tilde\rho \right)
\nonumber\\
&+&\frac{\tilde\Gamma_{\uparrow}}{2}\left(2 \sigma_+ \tilde\rho
\sigma_-
-\tilde\rho\sigma_-\sigma_+ - \sigma_-\sigma_+\tilde\rho \right)
\nonumber\\
&+&\frac{\tilde\Gamma_{\varphi}^*}{2} \left(\sigma_z\tilde\rho
\sigma_z -
\tilde\rho\right)\ ,
\end{eqnarray}
while the oscillator damping term is not affected by the transformation.
Although at low temperatures in the lab frame  the relaxation
processes dominate
(see Eq.~\ref{eq:Gamma_0}), the transformation to the rotating frame
introduces excitation and pure dephasing processes.

\section{The single-qubit laser}

In the following we will consider two resonance situations, $\Omega_R
\sim \omega_T$ or
$\Omega_R \sim 2\omega_T$, when either the one- or the two-photon
interactions dominate, and investigate the effects of blue or red
detuning, $\delta\omega\equiv\omega_{d}-\Delta E/\hbar$, of the
qubit driving frequency. We also study the effects of detuning of
the Rabi frequency $\Omega_R$ relative to that of the oscillator.

\subsection{One-photon interaction}

When the Rabi frequency is in resonance with the oscillator, $\Omega_R
\approx
\omega_T$, the Hamiltonian (\ref{V_I}) in RWA reduces to
\begin{eqnarray}\label{H_I1photon}
H_{I}&=&g_1\left(a^{\dagger}\sigma_-e^{-i(\Omega_R-\omega_T)
t}+h.c.\right)
\nonumber\\
&+&g_3\left(a^{\dagger}a + a a^{\dagger}\right)\sigma_z\ .
\end{eqnarray}
From here we can proceed in the frame of the standard semiclassical
approach~\cite{Reid,Gardiner} of laser physics with the following
main steps:
In the absences of fluctuations the system is described by
Maxwell-Bloch equations for the classical variables
$\alpha=\langle a \rangle$, $\alpha^*=\langle a^{\dagger}
\rangle$, $s_{\pm}=\langle \sigma_{\pm}\rangle$ and $s_z=\langle
\sigma_z\rangle$, which can be derived from the Hamiltonian
(\ref{H_I1photon}) if all correlation functions are assumed to
factorize. Next the qubit variables can be adiabatically
eliminated as long as $\kappa,\, g_1\ll\tilde\Gamma_1,\tilde\Gamma_
{\varphi}$,
which  leads to a closed equation of motion for $\alpha$. If we
finally account for fluctuations, e.g., due to thermal noise in
the resonator, $\alpha$ becomes a stochastic variable obeying
a Langevin equation~\cite{Gardiner},
\begin{eqnarray}\label{dot alpha}
\dot\alpha=-\Bigg[\kappa-\frac{C}{\tilde\Gamma_\varphi + i\delta \Omega}
s_z^{st}
+  4 i g_3 s_z^{st}\Bigg]\frac{\alpha}{2}+ \, \xi(t)
    \, .
\end{eqnarray}
Here
$C \equiv 2 g_1^2$,
$s_z^{st}=-D_0/\left(1+|\alpha^2|/\tilde n_0\right)$ is the
stationary value of the population difference between the qubit
levels, and
$D_0=\left(\tilde\Gamma_\downarrow-\tilde\Gamma_\uparrow\right)/\tilde
\Gamma_1$ is
the normalized difference between the rates with
$\tilde\Gamma_1=\tilde\Gamma_{\uparrow}+\tilde\Gamma_{\downarrow}$.
We further
introduced the photon saturation number
$n_0=\tilde\Gamma_{\varphi}\tilde\Gamma_1/4g_1^2$ and $\tilde
n_0 \equiv n_0(1+\delta\Omega^2/\tilde\Gamma_\varphi^2)$, and the total
dephasing
rate $\tilde\Gamma_\varphi=\tilde\Gamma_1/2+\tilde\Gamma_{\varphi}^*$.
The
detuning of the Rabi frequency enters in combination with a
frequency renormalization, $\delta\Omega \equiv \Omega_R-\omega_T
+ g_3 |\alpha|^2$. The Langevin force due to thermal noise in the
oscillator satisfies $\langle \xi(t)\xi^*(t')\rangle =
\kappa N_{\rm th}\delta(t-t')$ and $\langle \xi(t)\xi(t')\rangle = 0
$. Noise
originating from the qubit can be neglected provided the thermal
noise is strong, $\kappa N_{\rm th}\gg g_1^2/\tilde\Gamma_\varphi$.

\subsection{Two-photon interaction}

The two-photon effect
dominates near the resonance condition $\Omega_R\approx
2\omega_T$. In RWA the Hamiltonian reduces to
\begin{eqnarray}\label{H_I2photon}
H_{I}&=&g_2\left(a^{\dagger2}\sigma_-e^{-i(\Omega_R-2\omega_T)
t}+h.c.\right)\nonumber\\
&+&g_3\left(a^{\dagger}a+ a a^{\dagger}\right)\sigma_z \ .
\end{eqnarray}

The corresponding Langevin equation for the resonator variable
reads
\begin{eqnarray}\label{dot alpha2}
\dot\alpha=-\Bigg[\kappa-\frac{C}{\tilde\Gamma_\varphi + i\delta \Omega}
s_z^{st}
+  4 i g_3 s_z^{st}\Bigg]\frac{\alpha}{2}+ \, \xi(t)
    \, ,
\end{eqnarray}
i.e., is of the same form as Eq.~(\ref{dot alpha})
but with $C \equiv 4 g_2^2|\alpha|^2$  and $s_z^{st}=-D_0/\left(1+(|
\alpha^2|/\tilde n_0)^2\right)$. The
photon saturation number is now given by $n_0=(\tilde\Gamma_{\varphi}
\tilde\Gamma_1/4g_2^2)^{1/2}$,
and $\tilde n_0 \equiv n_0(1+\delta\Omega^2/\tilde\Gamma_\varphi^2)^
{1/2}$.
Again
$\xi(t)$ represents  thermal noise, while noise arising from
the qubit can be neglected if $\kappa N_{\rm th}\gg g_2^2 \bar
n/\tilde\Gamma_\varphi$. The detuning of the Rabi frequency for
two-photon
interaction is given by $\delta\Omega \equiv \Omega_R-2\omega_T +
g_3 |\alpha|^2$.
\begin{figure}
\includegraphics[width=0.9\columnwidth]
{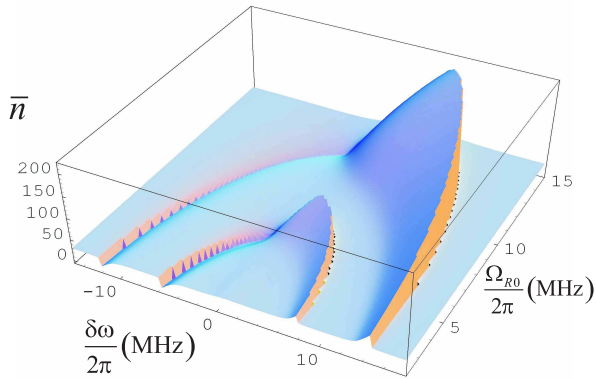} \caption{
Average number of photons in the resonator
as function of the driving detuning  $\delta\omega$ and amplitude
$\Omega_{R0}$. Peaks at $\delta\omega>0$ correspond to lasing,
while dips at $\delta\omega<0$ correspond to cooling. The inner
curve corresponds to the one-photon resonance which exists only
away from the symmetry point. Here we assumed $\epsilon =
0.01\Delta$. The outer curve describes the two-photon resonance,
which persists at $\epsilon = 0$. In domains of bistability the
lowest value of $\bar n$ is plotted (leading to the sharp drops in both curves). We chose the following
parameters for the qubit: $\Delta/2\pi = 1$~GHz,
$\epsilon=0.01\Delta$, $\Gamma_0/2\pi = 125$~kHz,
the resonator: $\omega_T/2\pi= 6$MHz, $\kappa/2\pi = 0.34$~kHz,
and the coupling: $g/2\pi=
3.3$~MHz. The bath temperature is $T=10$~mK.}
\label{3dphoton}
\end{figure}

\subsection{Results obtained from the Langevin equation}

If one neglects the frequency shifts of the oscillator, i.e., for
$g_3 = 0$, the Fokker-Planck equations corresponding to the
Langevin equations (\ref{dot alpha}) and (\ref{dot alpha2}) have
exact analytic solutions~\cite{Reid}. Also for
$g_3\neq0$ the Eqs.~(\ref{dot alpha}) and (\ref{dot alpha2}) written as
$\dot\alpha=- f(n)  \alpha/2 +\xi(t)$ can be transformed to equations
for the average number of photons $\langle|\alpha|^2\rangle=\bar
n$ in the form $\dot{\bar n}=-\langle n {\rm Re}[f(n)]\rangle +\kappa
N_{\rm th}$.
In the steady state, for $\bar n\gg 1$ they can be
approximated by $\bar n {\rm Re}[f(\bar n)]=\kappa N_{\rm th}$.
The results of this analysis are shown in Fig.~\ref{3dphoton}.
To demonstrate both
the one-photon and the two-photon effects we have assumed
a small deviation from the symmetry point, $\epsilon =
0.01\Delta$. The two-photon resonance (the outer one) persist
even for $\epsilon=0$, while the one-photon resonance (the inner one)
vanishes there. We observe that the solution shows bistability bifurcations
(see below). As a result we see in in Fig.~\ref{3dphoton}
sharp drops of $\bar n$ for both resonances as only the lowest
stable value is plotted .

We can estimate the asymptotic solutions analytically.
In the one-photon case, assuming $\bar n \gg \tilde n_0$, we obtain from Eq.Ö(\ref{dot alpha})
\begin{equation}\label{eq:n_av_saturated}
\bar n \sim N_{\rm th} + \frac{(-D_0)\tilde\Gamma_1}{2\kappa}\ .
\end{equation}
This result holds independent of whether the second contribution
due to the qubit is larger or smaller than the thermal number
$N_{\rm th}$ as long as $\bar n \gg \tilde n_0$. In the two-photon case,
assuming $\bar n \gg \tilde n_0$, we obtain from Eq.Ö(\ref{dot alpha2})
\begin{equation}\label{eq:n_av_saturated2}
\bar n \sim N_{\rm th} + \frac{(-D_0)\tilde\Gamma_1}{\kappa}\ .
\end{equation}

\subsection{Solution of the master equation}

\begin{figure}
\includegraphics[width=7cm]{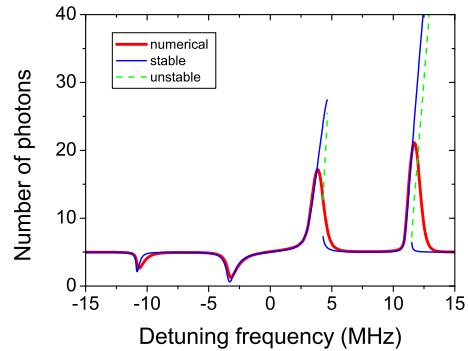}
\caption{Average number of photons $\bar n$ versus the detuning. The
blue curves are
obtained from the Langevin
equations (\ref{dot alpha}) and (\ref{dot alpha2}). They show the
bistability with the solid curve denoting stable solutions, while
the dashed curve denotes the unstable solution.
The red curve is obtained from a numerical solution of the master
equation (\ref{eq:Master_Equation}). The driving amplitude is taken
as $\Omega_{R0}/2\pi=5$MHz.
The parameters of the qubit: $\Delta/2\pi = 1$~GHz,
$\epsilon=0.01\Delta$, $\Gamma_0/2\pi = 125$~kHz,
the resonator: $\omega_T/2\pi= 6$MHz, $\kappa/2\pi = 1.7$~kHz,
$N_{\rm th}=5$, and the coupling: $g/2\pi=
3.3$~MHz. }
\label{fig:compar}
\end{figure}

We also solved the full master equation (\ref{eq:Master_Equation})
numerically, which provides  access not only to the average number of photons in
the oscillator, $\bar n$, but also to the whole distribution function $P
(n)$.
To reach convergence with a limited number of photon basis states
($ n \le 100$) we assumed a low thermal number, $N_{\rm th}=5$ and a relatively
high relaxation constant of the oscillator $\kappa/2\pi = 1.7$~kHz.
In Fig.~\ref{fig:compar} the solutions of the Langevin
equations (\ref{dot alpha}) and (\ref{dot alpha2}) and those of the
master equation (\ref{eq:Master_Equation}) is compared.

\begin{figure}
\includegraphics[width=7cm]{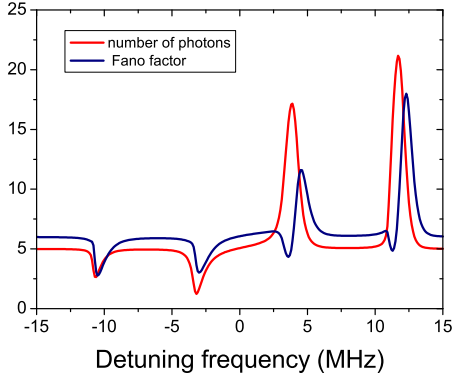}
\caption{The Fano factor (blue), and the average photon number $\bar n$
(red). The parameters are as in Fig.~\ref{fig:compar}.}
\label{fig:fano}
\end{figure}

In Fig.~\ref{fig:fano} the Fano factor
$F=\left( \langle n^2\rangle - \langle n\rangle^2\right)/\langle n
\rangle$ of the photon number distribution is presented.
We observe two phenomena. First, in the regime of the lasing
without bistability the Fano factor is reduced as compared to that of
the thermal state. In the bistable regime it is increased due to the switching between the two solutions.

\begin{figure}
\includegraphics[width=7cm]{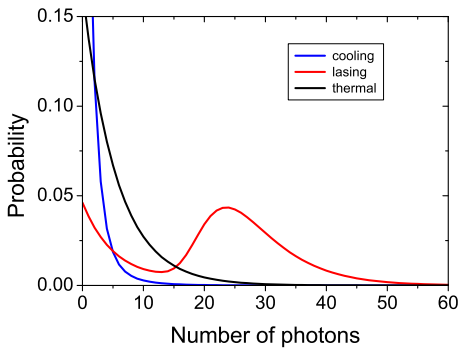}
\caption{
The distribution function, $P(n)$, obtained
by numerically solving the master equation (\ref{eq:Master_Equation}).
Blue curve: cooling regime of the one-photon resonance
with $\Omega_{R0} =2\pi \times 5$MHz and $\delta\omega = -2\pi \times3.2$MHz.
Red curve: lasing regime  of the two-photon resonance
with $\Omega_{R0} =2\pi \times5 $MHz $\delta\omega = 2\pi \times11.7$MHz.
We observe a peak in the $P(n)$ distribution between $n=20$ and 30 as a result of the lasing behavior.
Black curve: thermal distribution with $N_{\rm th}=5$. The parameters are as in Fig.~\ref{fig:compar}.
}
\label{fig:distributions}
\end{figure}

In Fig.~\ref{fig:distributions} the distribution function, $P(n)$,
for the number of photons in the oscillator is plotted both for the
cooling and enhancement regime. For
comparison also the thermal (Bose-Einstein) distribution is plotted.

\begin{figure}
\includegraphics[width=7cm]{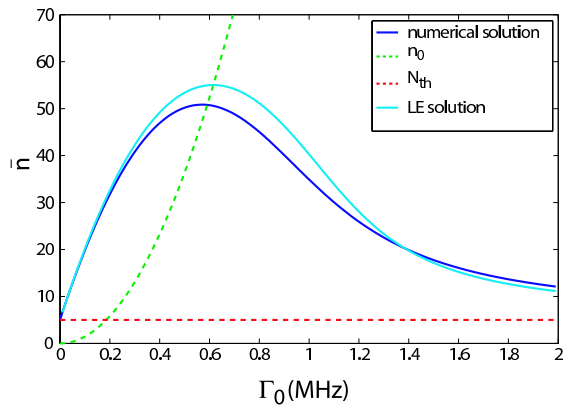}
\caption{Average number of photons in the resonator as function of
the qubit's
relaxation rate, $\Gamma_0$ at the one-photon resonance, $\Omega_R =
\omega_T$ for  $g_3=0$ and $N_{th}=5$. The  dark blue line shows the
numerical solution of the master equation, the light blue solid line
represents the solution of  the Langevin equation, Eq. (\ref{dot
alpha}). The green and red dashed curves represent respectively the
saturation number $n_0$ and the thermal photon number $N_{\rm th}$.
The parameters are as in Fig.~\ref{fig:compar} (except for $\Gamma_0$).}
\label{nvsga}
\end{figure}

The enhancement and cooling effects described here rely crucially on
the transition rates, i.e., the dissipative effects as described by
the Liouville equations. In order to demonstrate this dependence we
show in Fig.~\ref{nvsga} the dependence of the average
photon number on the qubit's relaxation rate at
the one-photon resonance. We note a non-monotonic dependence on the
qubit's relaxation rate. Above the saturation threshold for $\bar{n}>
n_0$  the pumping rate is limited by  $\Gamma_0$, leading to a
roughly linear growth of the photon number with increasing $\Gamma_0$,
consistent with Eq. (\ref{eq:n_av_saturated}). At the saturation threshold for $\bar n \sim
n_0$ the effective coupling is determined by $g_1$ and the photon
number becomes insensitive to small variations of $\Gamma_0$.
Finally, for $\bar n < n_0$, an increase of $\Gamma_0$ predominantly
increases the  dephasing rate $\tilde\Gamma_{\varphi}$.
As can be seen from Eq. (\ref{dot alpha}) this destroys the
coherent coupling between qubit and oscillator and the photon number decreases towards $N_{th}$. In Figure~\ref{nvsga} we plot both the results of a numerical solution of the
master equation  and the Langevin approximation. We find good
agreement between both (except for the bistability in the
corresponding parameter regions).

\section{Results and Discussion}

We  summarize our main conclusions.
Our results for the number of photons $\bar n$ are plotted in
Fig.~\ref{3dphoton} as a function of the detuning $\delta\omega$
of the driving frequency and driving amplitude $\Omega_{R0}$.
It exhibits sharp structures along two curves corresponding to the
one- and two-photon resonance conditions,
$\Omega_{R}=\omega_T-g_3\bar n$ and $\Omega_{R}=2\omega_T-g_3\bar
n$. Blue detuning, $\delta\omega>0$, induces a strong population
inversion of the qubit levels, which in resonance leads to
one-qubit lasing. In experiments the effect can be measured as a
strong increase of the number of photons in the resonator
above the thermal values. Red detuning
produces a one-qubit cooler with resulting photon
numbers substantially below the thermal value.

The bistability of the solution of the Langevin description is
illustrated in Fig.~\ref{fig:compar}. In the range of
bistability we expect a telegraph-like
noise corresponding to the random switches between the two
solutions.

Potentially useful applications of the considered scheme
are the lasing behavior and the creation of a highly non-thermal population of the oscillator as well as the cooling.
Within the accuracy of our approach we
estimate that a population of order $\bar n=1$
can be reached for optimal detuning. A more detailed analysis is
required to determine the precise cooling limit.

So far we described an LC oscillator coupled to a flux qubit.
But our analysis equally applies for a nano-mechanical
resonator coupled capacitively to a Josephson charge qubit (see
Fig.~\ref{fig:system}b). In this case $\sigma_z$ stands for the
charge of the qubit and both the coupling to the oscillator as well as
the driving are capacitive, i.e., involve $\sigma_z$. To produce the
capacitive coupling between the qubit and the oscillator, the
latter could be metal-coated and  charged by the voltage source $V_x$.
The dc component of the gate voltage $V_g$ puts the system near
the charge degeneracy point where the dephasing due to the $1/f$
charge noise is minimal. Rabi driving is induced by an $ac$
component of $V_g$. Realistic experimental parameters are expected
to be very similar to the ones used in the examples discussed
above, except that a much higher quality factor of the
resonator ($\sim 10^5$)
and a much higher number of quanta in the oscillator can be reached.
This number will easily exceed the thermal one,
thus a proper lasing state with Poisson statistics, appropriately named
SASER~\cite{SASER}, is produced.
One should then observe the usual  line narrowing with
line width  given by $\kappa N_{\rm th}/(4\bar n) \sim \kappa^2
N_{\rm th}/\tilde\Gamma_1$. Experimental observation of this line-width
narrowing would constitute a confirmation of the lasing/sasing.

\section{Acknowledgment}

We thank E. Il'ichev, O. Astafiev, A. Blais, M. Devoret, D. Esteve, M.~D. LaHaye, M. Marthaler, Y. Nakamura, F. Nori, E. Solano, K.~C. Schwab, and F.~K. Wilhelm
for fruitful discussions. The work is part of the EU IST Project EuroSQIP.

\bibliography{ref}

\begin{thebibliography}{10}

\bibitem{Jena_Rabi}
E.~Il\char39{}ichev, N.~Oukhanski, A.~Izmalkov, Th. Wagner, M.~Grajcar, H.-G.
  Meyer, A.~Yu. Smirnov, Alec Maassen van~den Brink, M.~H.~S. Amin, and A.~M.
  Zagoskin.
\newblock Continuous monitoring of {R}abi oscillations in a {J}osephson flux
  qubit.
\newblock {\em Phys. Rev. Lett.}, 91:097906, 2003.

\bibitem{Wallraff_CQED}
A.~Wallraff, D.~I. Schuster, A.~Blais, L.~Frunzio, R.-S. Huang, J.~Majer,
  S.~Kumar, S.~M. Girvin, and R.~J. Schoelkopf.
\newblock Circuit quantum electrodynamics: Coherent coupling of a single photon
  to a {C}ooper pair box.
\newblock {\em Nature}, 431:162, 2004.

\bibitem{Chiorescu_CQED}
I.~Chiorescu, P.~Bertet, K.~Semba, Y.~Nakamura, C.~J. P.~M. Harmans, and J.~E.
  Mooij.
\newblock Coherent dynamics of a flux qubit coupled to a harmonic oscillator.
\newblock {\em Nature}, 431:159, 2004.

\bibitem{wallraff-2005-95}
A.~Wallraff, D.~I. Schuster, A.~Blais, L.~Frunzio, J.~Majer, M.~H. Devoret,
  S.~M. Girvin, and R.~J. Schoelkopf.
\newblock Approaching unit visibility for control of a superconducting qubit
  with dispersive readout.
\newblock {\em Phys. Rev. Lett.}, 95:060501, 2005.

\bibitem{PhysRevLett.96.127006}
J.~Johansson, S.~Saito, T.~Meno, H.~Nakano, M.~Ueda, K.~Semba, and
  H.~Takayanagi.
\newblock Vacuum {R}abi oscillations in a macroscopic superconducting qubit
  {$LC$} oscillator system.
\newblock {\em Phys. Rev. Lett.}, 96(12):127006, Mar 2006.

\bibitem{naik-2006-443}
A.~Naik, O.~Buu, M.~D. LaHaye, A.~D. Armour, A.~A. Clerk, M.~P. Blencowe, and
  K.~C. Schwab.
\newblock Cooling a nanomechanical resonator with quantum back-action.
\newblock {\em Nature}, 443:193, 2006.

\bibitem{grajcar-2007}
M.~Grajcar, S.~H.~W. {van der Ploeg}, A.~Izmalkov, E.~Il'ichev, H.~G. Meyer,
  A.~Fedorov, A.~Shnirman, and G.~Sch\"on.
\newblock Sisyphus damping and amplification by a superconducting qubit.
\newblock {\em arXiv.org:0708.0665}, 2007.

\bibitem{deppe-2008}
F.~Deppe, M.~Mariantoni, E.~P. Menzel, A.~Marx, S.~Saito, K.~Kakuyanagi,
  H.~Tanaka, T.~Meno, K.~Semba, H.~Takayanagi, E.~Solano, and R.~Gross.
\newblock Two-photon probe of the {J}aynes-{C}ummings model and symmetry
  breaking in circuit {QED}.
\newblock {\em arXiv.org:0805.3294}, 2008.

\bibitem{Buisson_QED}
O.~Buisson, F.~Balestro, J.~P. Pekola, and F.~W.~J. Hekking.
\newblock One-shot quantum measurement using a hysteretic dc {SQUID}.
\newblock {\em Phys. Rev. Lett.}, 90:238304, 2003.

\bibitem{blais-2004-69}
A.~Blais, R.~S. Huang, A.~Wallraff, S.~M. Girvin, and R.~J. Schoelkopf.
\newblock Cavity quantum electrodynamics for superconducting electrical
  circuits: an architecture for quantum computation.
\newblock {\em Phys. Rev. A}, 69:062320, 2004.

\bibitem{liu-2004-67}
{Yu-xi} Liu, L.~F. Wei, and F.~Nori.
\newblock Generation of nonclassical photon states using a superconducting
  qubit in a microcavity.
\newblock {\em Europhys. Lett.}, 67:941, 2004.

\bibitem{martin-2004-69}
I.~Martin, A.~Shnirman, L.~Tian, and P.~Zoller.
\newblock Ground state cooling of mechanical resonators.
\newblock {\em Phys. Rev. B}, 69:125339, 2004.

\bibitem{moon:140504}
K.~Moon and S.~M. Girvin.
\newblock Theory of microwave parametric down-conversion and squeezing using
  circuit {QED}.
\newblock {\em Phys. Rev. Lett.}, 95:140504, 2005.

\bibitem{mariantoni-2005}
M.~Mariantoni, M.~J. Storcz, F.~K. Wilhelm, W.~D. Oliver, A.~Emmert, A.~Marx,
  R.~Gross, H.~Christ, and E.~Solano.
\newblock On-chip microwave fock states and quantum homodyne measurements.
\newblock {\em cond-mat/0509737}, 2005.

\bibitem{liu-2006-74}
{Yu-xi} Liu, C.~P. Sun, and F.~Nori.
\newblock Scalable superconducting qubit circuits using dressed states.
\newblock {\em Phys. Rev. A}, 74:052321, 2006.

\bibitem{wallquist-2006}
M.~Wallquist, V.~S. Shumeiko, and G.~Wendin.
\newblock Selective coupling of superconducting qubits via tunable stripline
  cavity.
\newblock {\em Phys. Rev. B}, 74:224506, 2006.

\bibitem{xue-2006}
Fei Xue, Y.~D. Wang, C.~P. Sun, H.~Okamoto, H.~Yamaguchi, and K.~Semba.
\newblock Controllable coupling between flux qubit and nanomechanical resonator
  by magnetic field.
\newblock {\em New J. Phys.}, 9:35, 2007.

\bibitem{Hauss_PRL2008}
J.~Hauss, A.~Fedorov, C.~Hutter, A.~Shnirman, and G.~Sch{\"o}n.
\newblock Single-qubit lasing and cooling at the {R}abi frequency.
\newblock {\em Phys. Rev. Lett.}, 100:037003, 2008.

\bibitem{ashhab-2008}
S.~Ashhab, J.~R. Johansson, A.~M. Zagoskin, and F.~Nori.
\newblock Single-artificial-atom lasing and its suppression by strong pumping.
\newblock {\em arXiv.org:0803.1209}, 2008.

\bibitem{wallquist-2008}
K.~Jaehne, K.~Hammerer, and M.~Wallquist.
\newblock Ground state cooling of a nanomechanical resonator via a {C}ooper
  pair box qubit.
\newblock {\em arXiv.org:0804.0603}, 2008.

\bibitem{Smirnov}
A.~\mbox{Yu.} Smirnov.
\newblock Theory of weak continuous measurements in a strongly driven quantum
  bit.
\newblock {\em Phys. Rev. B}, 68:134514, 2003.

\bibitem{greenberg-2005-72}
\mbox{Ya.} S.~Greenberg, E.~Il'ichev, and A.~Izmalkov.
\newblock Low frequency {R}abi spectroscopy for a dissipative two-level system.
\newblock {\em Europhys. Lett.}, 72:880, 2005.

\bibitem{PhysRevA.46.5944}
Yi~Mu and C.~M. Savage.
\newblock One-atom lasers.
\newblock {\em Phys. Rev. A}, 46(9):5944--5954, Nov 1992.

\bibitem{mckeever-2003-425}
J.~McKeever, A.~Boca, A.~D. Boozer, J.~R. Buck, and H.~J. Kimble.
\newblock Experimental realization of a one-atom laser in the regime of strong
  coupling.
\newblock {\em Nature}, 425:268, 2003.

\bibitem{Zakrewski}
J.~Zakrzewski, M.~Lewenstein, and T.~W. Mossberg.
\newblock Theory of dressed-state lasers. {I}. {E}ffective {H}amiltonians and
  stability properties.
\newblock {\em Phys. Rev. A}, 44:7717, 1991.

\bibitem{PhysRev.188.1969}
B.~R. Mollow.
\newblock Power spectrum of light scattered by two-level systems.
\newblock {\em Phys. Rev.}, 188(5):1969--1975, Dec 1969.

\bibitem{Plenio}
D.~Jonathan and M.~B. Plenio.
\newblock Light-shift-induced quantum gates for ions in thermal motion.
\newblock {\em Phys. Rev. Lett.}, 87(12):127901, Sep 2001.

\bibitem{SASER}
A.~J. Kent, R.~N. Kini, N.~M. Stanton, M.~Henini, B.~A. Glavin, V.~A. Kochelap,
  and T.~L. Linnik.
\newblock Acoustic phonon emission from a weakly coupled superlattice under
  vertical electron transport: Observation of phonon resonance.
\newblock {\em Phys. Rev. Lett.}, 96:215504, 2006.

\bibitem{Fulton1989}
T.~A. Fulton, P.~L. Gammel, D.~J. Bishop, L.~N. Dunkleberger, and G.~J. Dolan.
\newblock Observation of combined {J}osephson and charging effects in small
  tunnel junction circuits.
\newblock {\em Phys. Rev. Lett.}, 63:1307, 1989.

\bibitem{Geerligs}
A.~{Maassen van den Brink}, G.~Sch{\"o}n, and L.~J. Geerligs.
\newblock Combined single electron and coherent {C}ooper pair tunneling in
  voltage-biased {J}osephson junctions.
\newblock {\em Phys. Rev. Lett.}, 67:3030, 1991.

\bibitem{Maassen1991b}
A.~{Maassen van den Brink}, A.~A. Odintsov, P.~A. Bobbert, and G.~Sch{\"o}n.
\newblock Coherent {C}ooper pair tunneling in systems of {J}osephson junctions:
  Effects of quasiparticle tunneling and of the electromagnetic environment.
\newblock {\em Z. Physik B}, 85:459, 1991.

\bibitem{blanter-2004-93}
\mbox{Ya.} M.~Blanter, O.~Usmani, and \mbox{Yu.} V.~Nazarov.
\newblock Single-electron tunneling with strong mechanical feedback.
\newblock {\em Phys. Rev. Lett.}, 93:136802, 2004.

\bibitem{blencowe-2005-7}
M.~P. Blencowe, J.~Imbers, and A.~D. Armour.
\newblock Dynamics of a nanomechanical resonator coupled to a superconducting
  single-electron transistor.
\newblock {\em New J. Phys.}, 7:236, 2005.

\bibitem{clerk-2005-7}
A.~A. Clerk and S.~Bennett.
\newblock Quantum nano-electromechanics with electrons, quasiparticles and
  {C}ooper pairs: effective bath descriptions and strong feedback effects.
\newblock {\em New J. Phys.}, 7:238, 2005.

\bibitem{bennett-2006-74}
S.~D. Bennett and A.~A. Clerk.
\newblock Laser-like instabilities in quantum nano-electromechanical systems.
\newblock {\em Phys. Rev. B}, 74:201301, 2006.

\bibitem{usmani-2007}
O.~Usmani, Ya.~M. Blanter, and Yu.~V. Nazarov.
\newblock Strong feedback and current noise in nanoelectromechanical systems.
\newblock {\em Phys. Rev. B}, 75(19):195312, 2007.

\bibitem{rodrigues-2006}
D.~A. Rodrigues, J.~Imbers, and A.~D. Armour.
\newblock Quantum dynamics of a resonator driven by a superconducting
  single-electron transistor: A solid-state analogue of the micromaser.
\newblock {\em Phys. Rev. Lett.}, 98(6):067204, 2007.

\bibitem{Astafiev07}
O.~Astafiev, K.~Inomata, A.~O. Niskanen, T.~Yamamoto, Yu.~A. Pashkin,
  Y.~Nakamura, and J.~S. Tsai.
\newblock Single artificial-atom laser.
\newblock {\em Nature}, 449:588--590, 2007.

\bibitem{greenberg-2002-66}
Ya.~S. Greenberg, A.~Izmalkov, M.~Grajcar, E.~Il\char39{}ichev, W.~Krech, H.-G.
  Meyer, M.~H.~S. Amin, and A.~Maassen van~den Brink.
\newblock Low-frequency characterization of quantum tunneling in flux qubits.
\newblock {\em Phys. Rev. B}, 66:214525, 2002.

\bibitem{Saclay_Karlsruhe}
G.~Ithier, E.~Collin, P.~Joyez, P.J. Meeson, D.~Vion, D.~Esteve, F.~Chiarello,
  A.~Shnirman, Yu. Makhlin, J.~Schriefl, and G.~Sch{\"o}n.
\newblock Decoherence in a superconducting quantum bit circuit.
\newblock {\em Phys. Rev. B}, 72:134519, 2005.

\bibitem{Gardiner}
C.~W. Gardiner and P.~Zoller.
\newblock {\em Quantum noise}.
\newblock Springer, 3-d edition, 2004.

\bibitem{Reid}
M.~Reid, K.~J. McNeil, and D.~F. Walls.
\newblock Unified approach to multiphoton lasers and multiphoton bistability.
\newblock {\em Phys. Rev. A}, 24:2029, 1981.

\end{thebibliography}

\end{document}